\documentclass[12pt]{article}
\usepackage{graphicx}
\usepackage{epstopdf}
\usepackage{amssymb,amsmath,epsfig}

\begin{document}

\title{\bf Charged Cylindrical Polytropes with Generalized Polytropic Equation of State}

\author{M. Azam$^{1}$ \thanks{azam.math@ue.edu.pk, azammath@gmail.com},
S. A. Mardan$^2$ \thanks{syedalimardanazmi@yahoo.com, ali.azmi@umt.edu.pk},
I. Noureen$^2$ \thanks{ifra.noureen@gmail.com, ifra.noureen@umt.edu.pk} and
M. A. Rehman$^2$ \thanks{aziz3037@yahoo.com, aziz.rehman@umt.edu.pk}\\
$^1$ Division of Science and Technology, University of Education,\\
Township Campus, Lahore-54590, Pakistan.\\
$^2$ Department of Mathematics,\\ University of the Management and Technology,\\
C-II, Johar Town, Lahore-54590, Pakistan.}

\date{}

\maketitle
\begin{abstract}
We study the general formalism of polytropes in relativistic regime with  generalized
polytropic equations of state in the vicinity of cylindrical symmetry.
We take charged anisotropic fluid distribution of matter with conformally
flat condition for the development of general framework of polytropes.
We discussed the stability of the model by Whittaker formula and concluded
that one of the developed model is physically viable.
\end{abstract}
{\bf Keywords:} Relativistic Anisotropic Fluids; Polytropes; Electromagnetic Field; Cylindrical symmetry.\\

\section{Introduction}

In general relativity (GR), polytropes play very vital role in the modeling of relativistic
compact objects (CO). Over the past few decades,
many researchers have been engaged in the study of polytropes
due to simple form of polytropic equation of state (EoS) and
corresponding Lane-Emden equation (LEE) that can be used for the description
of various astrophysical phenomenons. Chandrasekhar \cite{1} was the pioneer, who established the theory of
polytropes originated through laws of thermodynamics in Newtonian regime. Tooper \cite{2,3} formulated
the initial framework of polytropes for compressible and adiabatic fluid
under quasi-static equilibrium condition to develop LEE. After few years,
Kovetz \cite{4} provided some corrections in Chandrasekhar formalism for polytropes.
The general form of LEE in higher dimensional space was developed by
Abramowicz \cite{5} in spherical, cylindrical and planner geometry.

The study of anisotropy in the modeling of astrophysical CO
plays a very significant role and many physical problem cannot be
modeled without taking account anisotropic stress. In 1981, a very
sophisticated process of modeling anisotropic CO was provided by Cosenza \cite{6}.
In this scenario, Herrera and Santos \cite{7} presented a detailed study
for the existence of anisotropy in self-gravitating CO. Herrera and Barreto \cite{8} developed a general
model for Newtonian and relativistic anisotropic polytropes. Herrera et al. \cite{9}
gave a complete set of equations with anisotropic stress for
self-gravitating spherical CO. A new way to check the stability
of anisotropic polytropic models through Tolman mass was provided by
Herrera and Barreto \cite{10,11}. Herrera et al. \cite{12} used
conformally flat condition in the analysis of anisotropic polytropes
to reduce parameters involve in LEE. He and his coworkers \cite{13}
also discussed the implementation of cracking criteria for the
stability of anisotropic polytropes.

In GR, the existence of charge considerably affect the modeling of relativistic CO.
Bekenstein \cite{14} investigated gravitational collapse by means
of hydrostatic equilibrium equation in charged CO.
Bonnor \cite{15,16} showed that gravitational collapse
can be delayed by electric repulsions in CO. A complete study of the contraction
of charged CO in isotropic coordinates was presented by Bondi \cite{17}.
Koppar et al. \cite{18} developed a novel way to calculate
charge generalization of static charged fluid solution of CO.
Ray et al. \cite{19} investigated charged CO with high density and found that
they can have large amount of charge approximately $10^{20}$ coulomb.
Herrera et al. \cite{20} used structure scalars to study dissipative
fluids in charged spherical CO. Takisa \cite{21} provided models of
polytropic CO in the presence of charge.
Sharif \cite{22} developed modified LEE for charged polytropic CO with
conformally flat condition.
Azam et al. \cite{23} discussed the cracking of
different charged CO models with liner and quadratic EoS.

It is always a crucial issue to choose proper EoS
for the modeling of astronomical objects. Chavanis proposed
a modification in conventional polytropic EoS
$P_r=K \rho^{1+\frac{1}{n}}$, where $P_r$ is radial pressure,
$n$ is polytropic index and $K$ is polytropic constant.
He combined linear EoS $P_r=\alpha_1 \rho_o$ with polytropic EoS
as $P_r=\alpha_1 \rho_o+K \rho^{1+\frac{1}{n}}$, and used
it to describe different cosmological situations. He
developed the models of early and late universe
for $n>0$ and $n<0$ through generalized polytropic
equation of state (GPEoS) \cite{28}.
Freitas \cite{30} applied the modified polytropic EoS
for the development of universe model with constant energy density
and discussed quantum fluctuations of universe.
Azam et al. \cite{31} provided a comprehensive study for the development
of modified form of LEE with GPEoS for spherical symmetry.

Cylindrically symmetric spacetimes have been used widely in GR to describe various
physically interesting aspects. For the first time, Kompaneets \cite{32}
provided the general form of four-dimensional cylindrically symmetric metric.
Some specific examples of cylindrically symmetric metric which provide exact solutions
to the system of Einstein's field equations and cylindrical gravitational waves were studied in \cite{33,34}.
Thorne \cite{34} defined the C-energy of cylindrical systems defined as
``gravitational energy per unit specific length". The most interesting fact about
these spacetimes is the so-called C-energy and as a result
gravitational waves are thought to be carrier of energy in
gravitational field. Whittaker \cite{35} introduced the concept of "mass potential"
in GR. Herrera et al. \cite{36} used conformally flat condition with
cylindrical symmetry to gave solution of field equations which is completely
matched to Levi-Civita vacuum spacetime. Herrera and Santos \cite{37}
studied matching condition for perfect fluid cylindrical gravitational collapse.
Debbasch et al. \cite{38} discussed regularity and matching condition for
stationary cylindrical anisotropic fluid. Di Prisco et al. \cite{39}
studied cylindrical gravitational collapse with shear-free condition.
Sharif and Fatima \cite{40} presented cylindrical collapse with charged
anisotropic fluid. Sharif and Azam \cite{41} studied dynamical instability of cylindrical
collapse in Newtonian and post Newtonian regime.
Ghua and Banerji \cite{42} described dissipative cylindrical collapse
with charge anisotropic fluid. Sharif and Sadiq \cite{43}
presented conformally flat polytropes with anisotropic fluid
for cylindrical geometry. Mahmood et al. \cite{44} considered
charged anisotropic fluid for the discussion of cylindrical collapse
and found that presence of charge enhances the anisotropy
of the collapsing system.

In this paper, we will explore charged anisotropic polytropes
by using GPEoS for a cylindrical symmetric
configuration with conformally flat condition.
In Sect. \textbf{2}, we present
the Einstein-Maxwell field equations and modified
hydrostatic equilibrium equation.
In Sect. \textbf{3}, LEE is developed for relativistic polytropes.
The energy conditions, conformally flat condition
and stability of the model is given in Sect. \textbf{4}.
In the last section we conclude our results.

\section{Matter Distribution and Einstein-Maxwell Field Equations}

In this section, we will describe the inner matter distribution
and Einstein-Maxwell's field equations.
We assume static cylindrically symmetric spacetime
\begin{equation}\label{1}
ds^2=-A^{2}dt^{2}+B^{2}dr^{2}+C^2 d\theta^{2}+{d z^2},
\end{equation}
where $t\in(-\infty,\infty)$, $r\in[0,\infty)$,
$\theta\in[0,2\pi]$ and $z\in(-\infty,\infty)$ are the
conditions on the cylindrical coordinates.
The energy-momentum tensor for charged anisotropic fluid distribution is
\begin{eqnarray}\label{2}\notag
T_{i j}&=&(P_r+\rho) V_{i} V_{j} -(P_r-P_z)S_{i}S_{j}
 -(P_r-P_\theta)K_{i}K_{j}+P_r g_{ij}\\
&+&\frac{1}{4\pi}(F_i^\gamma F_{j \gamma}-
\frac{1}{4}F^{\gamma\beta} F_{\gamma\beta}g_{ij} ),
\end{eqnarray}
where $P_r,~P_\theta,~P_z$ and $\rho$ represent
pressures in $r,~\theta,~z $ directions and energy density of
fluid inside cylindrical symmetric distribution. The four velocity
$V_{i}$ and four vectors $S_{i},~K_{i}$ satisfying the following relations
\begin{equation}\label{3}
V^{i}V_{i} = -1,~S^{i}S_{i}=K^{i}K_{i}=1,~S^{i}k_{i}=V^{i}k_{i}=V^{i}S_{i}=0.
\end{equation}
These quantities in co-moving coordinates can be written as
\begin{eqnarray}\label{4}
V_{i}=A\delta_i^0,~K_{i}=C\delta_i^2,~S_{i}=A\delta_i^3.
\end{eqnarray}
The Maxwell field equations are
\begin{equation}\label{5}
F_{[ij;k]}=0,~~~ F^{ij}_{;j}=4\pi J^i,
\end{equation}
where $F_{ij}=\psi_{j,i}-\psi_{i,j}$ is field tensor and
$\psi_i$ is the four-potential and  $J^i$ is four-current.
The four-potential and four-velocity are related to each other
in co-moving coordinates as
\begin{equation}\label{6}
\psi_{i}=\psi(r) \delta^0_i,~~~J^i=\sigma V^i,~~~~~~ i=0,1,2,3,
\end{equation}
with $\psi$ and $\sigma$ represents scalar potential  and charge
density respectively.\\
The Einstein-Maxwell field equations for line element Eq.$(\ref{1})$
are given by
\begin{eqnarray}
&&\frac{B^\prime C^\prime}{B^3 C}-\frac{C^{\prime\prime}}{B^2C}=
8\pi \rho -4\pi^2 E^2, \label{7}\\
&&\frac{A^\prime C^\prime}{A B^2 C}=8\pi P_r +4\pi^2 E^2, \label{8}\\
&&\frac{A^{\prime\prime}}{A B^2}-\frac{A^\prime B^\prime}{A B^3}=
8\pi P_\theta -4\pi^2 E^2, \label{9}\\
&&\frac{A^{\prime\prime}}{A B^2}-\frac{A^\prime B^\prime}{A B^3}
-\frac{B^\prime C^\prime}{B^3 C}-\frac{A^\prime C^\prime}{A B^2 C}
+\frac{C^{\prime \prime}}{ B^2 C}=8\pi P_z -4\pi^2 E^2, \label{10}
\end{eqnarray}
where $``\prime"$ denotes the differentiation with respect to $r$ and
$E=\frac{q}{2\pi C}$ with $q(r)=4 \pi\int_0^r \sigma B C dr$ represents total
amount of charge per unit length of cylinder.
We consider the exterior metric for the cylindrical symmetric geometry
with retarded time coordinate $\nu$ defined as
\begin{equation}\label{11}
ds^2=-(-\frac{2M}{R}+\frac{Q^2}{R^2})d\nu^2-2d\nu dR+R^2(d\theta^2+\beta^2dz^2),
\end{equation}
where $M$ is the total mass and $\beta$ is arbitrary constant.\\
The junction condition has a very important role
in the theory of relativistic objects.
These condition provide us the feasibility of
physically acceptable solutions.
For the continuity and matching of two spacetimes,
junction conditions  on the boundary $\Sigma$ yields \cite{45,46}
\begin{equation}\label{12}
m(r)-M\overset\Sigma =\frac{l}{8},~~\Leftrightarrow
~~Q^2\overset\Sigma=\frac{q l^2}{8},~~l\overset\Sigma=4C,~~P_r\overset\Sigma=0.
\end{equation}
The Schwarzschild coordinate is selected as $C=r$ \cite{43} and
Einstein-Maxwell field equations reduced to the form \cite{44}
\begin{eqnarray}
&&\label{13}\frac{B^\prime}{r B^3}=8\pi \rho -4\pi^2 E^2, \\
&&\label{14}\frac{A^\prime }{A B^2 r}=8\pi P_r +4\pi^2 E^2, \\
&&\label{15}\frac{A^{\prime\prime}}{A B^2}-\frac{A^\prime B^\prime}{A B^3}
=8\pi P_\theta -4\pi^2 E^2, \\
&&\label{16}\frac{A^{\prime\prime}}{A B^2}-\frac{A^\prime B^\prime}{A B^3}
-\frac{B^\prime }{B^3 r}-\frac{A^\prime }{A B^2 r}
=8\pi P_z -4\pi^2 E^2.
\end{eqnarray}
Solving Eqs.$(\ref{13})$-$(\ref{15})$ simultaneously lead to
hydrostatic equilibrium equation
\begin{equation}\label{17}
\frac{d P_r}{dr}+\frac{A^\prime}{A}\big(\rho +P_r\big)+\frac{\Delta}{r}
+\pi E E^\prime+\frac{8\pi E^2}{r}=0,
\end{equation}
where we have used $\Delta=P_r-P_\theta$.\\
Thorne \cite{34} defined C-energy (gravitational energy per
unit specific length of cylindrical geometry), in the form of mass function
\begin{equation*}
\widetilde{E} =\frac{1}{8}[1-l^{-2}\nabla^a\widetilde{r}\nabla_a\widetilde{r}],
\end{equation*}
with
\begin{equation*}
\mu^2=\xi_{(1)a}\xi^a_{(1)},~~~l^2=\xi_{(2)a}\xi^a_{(2)},~~~\widetilde{r}=\mu l,
\end{equation*}
here $\widetilde{r},\mu,l$ represents areal radius,
circumference radius, specific length respectively and for static case
expression of C-energy can be written as
\begin{equation}\label{18}
m(r)=\widetilde{E}^\prime =\frac{1}{8}\Big[1-\frac{1}{B^2}\Big]+2\pi r^2 E,
\end{equation}
Differentiating Eq.$(\ref{18})$ and using Eq.$(\ref{14})$, we get
\begin{equation}\label{19}
\frac{A^\prime}{A}=\frac{8\pi r P_r+4\pi^2r E^2}{1-8m+16\pi^2rE^2}.
\end{equation}
Using Eq.$(\ref{19})$, the hydrostatic equilibrium equation $(\ref{17})$ become
\begin{equation}\label{20}
\frac{d P_r}{dr}+\frac{8\pi r P_r+4\pi^2r E^2}{1-8m+16\pi^2rE^2}
\big(\rho +P_r\big)+\frac{\Delta}{r}
+\pi E E^\prime+\frac{8\pi E^2}{r}=0.
\end{equation}
The basic theory of polytropes is established
with the hypothesis of polytropic EoS
and hydrostatic equilibrium state of the
relativistic object under consideration.
In the next section, we will discuss relativistic
polytropes with generalize polytropic EoS
in the presence of charge for cylindrical symmetry.

\section{The Relativistic Polytropes}

In this section, we provide a comprehensive way for the development of
LEE which is the main consequence of theory of polytropes with
GPEoS in cylindrical regime. The EoS is the union
of linear EoS  $``P_r=\alpha_1\rho_o" $ and polytropic EoS
$``P_r=K\rho_o^{1+\frac{1}{n}}"$. The linear EoS describes
pressureless $(\alpha_1=0)$ or radiation $(\alpha_1=\frac{1}{3})$ matter.
The polytropic part is related to the cosmological aspects of early
universe for ``$n>0$'' whereas it describes late universe with ``$n<0$'' \cite{18,19}.
The cosmic behavior of universe is demonstrated with $``\rho_o"$ as
Planck density but in relativistic regime we take it as mass density for case \textbf{1} and
total energy density in case \textbf{2}.
Here, we shall develop the general formalism for relativistic polytropes with GPEoS
in the presence of charge.

\subsection{Case 1}

Here, the GPEoS is
\begin{equation}\label{21}
P_r=\alpha_1\rho_o+K\rho_o^{\gamma}=\alpha_1\rho_o+K\rho_o^{1+\frac{1}{n}}.
\end{equation}
The original polytropic part remain conserved
and the relationship of mass density $\rho_0$ with total energy density
$\rho$ is given by \cite{7}
\begin{equation}\label{22}
\rho=\rho_{o}+n P_r.
\end{equation}
Now taking following assumptions
\begin{equation}\label{23}
\alpha=\frac{P_{rc}}{\rho_{gc}},~~~r=\frac{\xi}{X},~~~\rho_{o}=\rho_{gc}\theta^n,~~~
m(r)=\frac{2\pi\rho_{gc} v({\xi})}{X^2},~~~X^2=\frac{16\pi P_{rc}}{\alpha},
\end{equation}
where $P_{rc}$  and $\rho_{gc}$ represents the central pressure and mass density.
Also $\xi$, $\theta$ and
$v$ defined to be dimensionless variables.
Using assumptions $(\ref{23})$ with EoS
$(\ref{21})$, the hydrostatic equilibrium equation Eq.$(\ref{20})$ transform as
\begin{eqnarray}\label{24} \notag
&&\Big(\frac{1-v({\xi})+4\pi^{3/2}\sqrt{\frac{\alpha}
{ P_{rc}}}E^2 \xi}{1+(n+1)(\alpha_1+\alpha\theta)}\Big)\frac{d \theta}{d \xi}+
\Big(\frac{2P_{rc}\alpha(\alpha_1+\alpha \theta)\theta^n+\pi E^2}
{n\alpha_1 \theta^{-n}+(n+1)\alpha\theta}\Big)\frac{\alpha\xi}{4P_{rc}}\\
&&+ \frac{(\Delta+\pi \xi E \frac{d E}{d \xi}+8 \pi E^2)(1-v({\xi})+4\pi^{3/2}
\sqrt{\frac{\alpha}{ P_{rc}}}E^2 \xi)}{(1+(n+1)(\alpha_1+\alpha\theta))
(n\alpha_1 \theta^{-n}+(n+1)\alpha)}\frac{\theta^n}{\alpha \xi P_{rc}}=0.
\end{eqnarray}
Now taking derivative of Eq. $(\ref{18})$ with respect to
$``r"$ and applying relations of Eq. $(\ref{23})$, we obtain
\begin{eqnarray}\label{25}
\frac{dv(\xi)}{d\xi}=\xi \theta^n(1+n\alpha_1+n\alpha\theta)
+\frac{(4-\pi)\alpha\xi E}{2 P_{rc}}+\frac{\alpha\xi^2}{ P_{rc}}\frac{d E}{d\xi}.
\end{eqnarray}
The combination of Eqs. $(\ref{24})$ and $(\ref{25})$
results the modified LE (Eq.$(\ref{501})$ in appendix),
which describe the relativistic charged polytropes with GPEoS.

\subsection{Case 2}

In this case the GPEoS as $P_r=\alpha_1\rho+K\rho^{1+\frac{1}{n}}$,
here mass density $\rho_o$ is replaced by
total energy density $\rho$ in Eq.$(\ref{21})$ and
have following relation \cite{7}
\begin{equation}\label{27}
\rho=\frac{\rho_o}{\big(1-K \rho_o^{\frac{1}{n}}\big)^n}.
\end{equation}
We take following assumptions
\begin{equation}\label{28}
\alpha=\frac{P_{rc}}{\rho_{c}},~~~
r=\frac{\xi}{X},~~~\rho=\rho_{c}\theta^n,~~~
m(r)=\frac{2\pi\rho_{c}v({\xi})}{X^2},~~~X^2=\frac{16\pi P_{rc}}{\alpha},
\end{equation}
where $c$ means that each quantity is calculated at center of CO.
Using GPEoS and assumptions of Eq. $(\ref{28})$,
the hydrostatic equilibrium equation $(\ref{20})$ turns out to be
\begin{eqnarray}\label{29} \notag
&&\Big(\frac{1-v({\xi})+4\pi^{3/2}\sqrt{\frac{\alpha}
{ P_{rc}}}E^2 \xi}{1+\alpha_1+\alpha\theta}\Big)\frac{d \theta}{d \xi}+
\Big(\frac{2P_{rc}\alpha(\alpha_1+\alpha \theta)\theta^n+\pi E^2}
{n\alpha_1 \theta^{-n}+(n+1)\alpha\theta}\Big)\frac{\alpha\xi}{4P_{rc}}\\
&&+ \frac{(\Delta+\pi \xi E \frac{d E}{d \xi}+8 \pi E^2)(1-v({\xi})+4\pi^{3/2}
\sqrt{\frac{\alpha}{ P_{rc}}}E^2 \xi)}{(1+\alpha_1+\alpha\theta)
(n\alpha_1 \theta^{-n}+(n+1)\alpha\theta)}\frac{\theta^n}{\alpha \xi P_{rc}}=0.
\end{eqnarray}
Taking derivative of Eq. $(\ref{18})$ with respect to
$``r"$ and applying Eq. $(\ref{28})$, we get
\begin{eqnarray}\label{30}
\frac{dv(\xi)}{d\xi}=\xi\theta^n+\frac{(4-\pi)\alpha\xi E}{2 P_{rc}}+\frac{\alpha\xi^2}{P_{rc}}\frac{d E}{d\xi}.
\end{eqnarray}
We get modified LEE by using Eqs. $(\ref{30})$ and Eq. $(\ref{29})$
given in (\textbf{Appendix} $(\ref{502})$), represents the relativistic charged polytropes with GPEoS.

\section{Energy Conditions, Conformally Flat Condition and Stability Analysis}

In the mathematical modeling of CO, the energy conditions plays
a very peculiar role in the analysis of developed model.
The energy condition provide us the maximum information
without depending upon EoS used in the modeling.
These conditions have been developed with understanding
that energy density is always positive otherwise the empty space
created due to positive and negative regions definitely become unstable.
The energy conditions should be satisfied by all the models are \cite{47}
\begin{eqnarray}\label{31}
\rho+\frac{\pi}{2}E^2>0,~~~P_{r} \leq \rho-\pi
E^2,~~~\frac{P_{\theta}}{\rho}\leq ,~~~\frac{P_{z}}{\rho}\leq 1.
\end{eqnarray}
If we take case \textbf{1} of developed model,
the energy conditions $(\ref{31})$ transform as
\begin{eqnarray}\notag
&&1+\frac{\pi E^2 }{1+n(\alpha_1 + \alpha\theta)\theta^{n}}>0,~~~
1\leq \frac{\pi E^2}{(n-1)(\alpha_1+\alpha\theta)},
\\&&\label{32}
\frac{\alpha P_\theta}{P_{rc}(1+n(\alpha_1+\alpha\theta))\theta^n}\leq 1,~~~
\frac{\alpha P_z}{P_{rc}(1+n(\alpha_1+\alpha\theta))\theta^n}\leq 1,
\end{eqnarray}
and for case \textbf{2} the energy conditions $(\ref{31})$ emerge as
\begin{eqnarray}\label{33}
1+\frac{\pi \alpha \theta^{-n} E^2}{2 P_{rc}}>0,
~1 \leq \frac{\pi\alpha E^2 \theta^n+\alpha P_{rc}
 \theta}{P_{rc}(1-\alpha_1)},~
\frac{P_\theta \alpha \theta^{-n}}{P_{rc}}\leq 1,
~\frac{P_z \alpha \theta^{-n}}{P_{rc}}\leq 1.
\end{eqnarray}
We observe that coupled Eqs.$(\ref{24})$-$(\ref{25})$
and Eqs.$(\ref{29})$-$(\ref{30})$ results a system
of differential equations with three variables.
Thus, we need more information to study charged polytropic
CO with GPEoS in cylindrical symmetry. So, we use
conformally flat condition to reduce a
systems of  differential equations to two variables.\\
For this purpose, the Weyl scalar defined in the form of Kretchman scalar,
Recci tensor and Recci scalar given by \cite{43}
\begin{eqnarray}\label{34}
C^2=\mathcal{R}+\frac{R^2}{3}-2R^{\mu\nu}R_{\mu\nu}.
\end{eqnarray}
For our line element, above equation become
\begin{eqnarray}\label{35}\notag
C^2&=&\frac{4}{3 r^2 A^2 B^6} \Big[\Big(A^2-r A A'
+r^2 A'^2\Big) B'^2+\notag\\&&
B^2 \Big(A'^2-r A' A''+r^2 A''^2\Big)+\notag\\&&
B B' \Big(r A' \Big(A'-2 r A''\Big)+A \Big(A'+r A''\Big)\Big)\Big].
\end{eqnarray}
Now applying conformally flatness i.e., $C^2=0$, and using field
equations $(\ref{13})$-$(\ref{16})$ and $(\ref{19})$
in the above Eq.$(\ref{35})$, we obtain anisotropy as
{
\allowdisplaybreaks
\begin{eqnarray}\label{36}\notag
\Delta &=&-\pi  E^2+\frac{1}{8 \pi }(8 \pi  P_r+4 \pi ^2 E^2+
\Big(8 \pi  P_r+4 \pi ^2 E^2\Big)^2\notag \\
&&+\Big(8 \pi  P_r+4 \pi ^2 E^2\Big) \Big(-4 \pi ^2 E^2+8
\pi  \rho \Big)+\Big(-4 \pi ^2 E^2+8 \pi  \rho \Big)^2\notag \\
&&+\Big(-1-8 \pi  P_r-8 \pi ^2 E^2+8 \pi  \rho \Big)
\Bigg[\frac{\Big(8 \pi  P_r+4 \pi ^2 r E^2\Big)^2}{\Big(1-8
m+16 \pi ^2 r E^2\Big)^2}\notag \\ &&-\frac{r
\Big(8 \pi  P_r+4 \pi ^2 r E^2\Big)
\Big(-4 \pi ^2 E^2+8 \pi  \rho \Big)}{1-8 m+16 \pi ^2 r E^2}+\notag \\ &&
\frac{4 \pi ^2 E^2+8 \pi  P_r'+8 \pi ^2 r E E'}
{1-8 m+16 \pi ^2 r E^2}-\notag \\ &&
 \frac{\Big(8 \pi  P_r+4 \pi ^2 r E^2\Big) \Big(16 \pi ^2 E^2-8 m'+
 32 \pi ^2 r E E'\Big)}{\Big(1-8 m+16 \pi ^2 r E^2\Big)^2}\Bigg]+\notag \\ &&
\Bigg[\frac{\Big(8 \pi  P_r+4 \pi ^2 r E^2\Big)^2}{\Big(1-8 m+16 \pi ^2
r E^2\Big)^2}-\frac{r \Big(8 \pi  P_r+4 \pi ^2 r E^2\Big)
\Big(-4 \pi ^2 E^2+8 \pi  \rho \Big)}{1-8 m+16 \pi ^2 r E^2}+ \notag \\ &&
\frac{4 \pi ^2 E^2+8 \pi  P_r'+8 \pi ^2 r E E'}{1-8 m+16 \pi ^2 r E^2}-\notag \\ &&
  \frac{\Big(8 \pi  P_r+4 \pi ^2 r E^2\Big)
  \Big(16 \pi ^2 E^2-8 m'+32 \pi ^2 r E E'\Big)}{\Big(1-8 m+16 \pi
^2 r E^2\Big)^2}\Big)^2\Bigg].
\end{eqnarray}
}
Now using Eq. $(\ref{23})$ in above equation,
we obtain anisotropy factor for case \textbf{1} given in (\textbf{Appendix} $(\ref{503})$).
One can derive modified LEE for conformally flat polytropes 
by using Eqs. $(\ref{503})$ in $(\ref{24})$ and coupling 
with Eq. $(\ref{25})$ for case \textbf{1}.
Similarly, anisotropy parameter for case \textbf{2} turn out to be
Eq. $(\ref{504})$ (see \textbf{Appendix}) and modified LEE by using Eqs. $(\ref{504})$
in $(\ref{29})$ and coupling it with Eq.$(\ref{30})$ for case \textbf{2},
which can help in the study of conformally flat polytropes.

We will use modify Whittaker \cite{35} formula for the
stability analysis of the model, which is the measure of
``active gravitational mass per unit length" of cylinder defined by
\begin{eqnarray}\label{37}
m_L = 8 \pi \int_0^{r_\Sigma} (\rho + P_r + P_\theta + P_z-\pi E^2)dr .
\end{eqnarray}
For case \textbf{1}, using field Eqs.$(\ref{13})$-$(\ref{16})$,
$(\ref{19})$ and $(\ref{23})$ in $(\ref{37})$, we get
{
\allowdisplaybreaks\begin{eqnarray}\label{38}\notag
&&m_L =\int_0^{1}-\frac{\pi\sqrt{(-\frac{2M}{R}+\frac{Q^2}{R^2})
\frac{P_{\text{rc}}}{\alpha }}}{\alpha  \Big(4 \pi ^{3/2}
\alpha  \xi  E^2 \sqrt{\frac{P_{\text{rc}}}{\alpha }}-
P_{\text{rc}} (-1+v)\Big) \theta }\notag \\ &&
  \Big(-16 \pi ^{5/2} \alpha ^2 \xi  E^3 \sqrt{\frac{P_{\text{rc}}}{\alpha }}
  \theta +\pi ^2 \alpha ^2 \xi  E^4 \Big(3 \xi
+32 \sqrt{\pi } \sqrt{\frac{P_{\text{rc}}}{\alpha }}\Big) \theta
+4 \pi  E P_{\text{rc}}  \notag \\ &&
\theta  \Big(-\alpha +2 \xi  \Big(-\alpha +32 \pi
P_{\text{rc}} \theta ^n \Big(\alpha _1+\alpha  \theta \Big)\Big)
\frac{d E}{d \xi} +v \Big(\alpha +2 \alpha  \xi
\frac{d E}{d \xi}\Big)\Big)+\notag \\ &&
8 P_{\text{rc}} \Big(-\alpha  \Delta  (-1+v) \theta +2
\sqrt{\pi } \alpha  \xi  \Big(\frac{P_{\text{rc}}}{\alpha }\Big){}^{3/2} \alpha
_1 \Big(1+n \alpha _1\Big) \theta ^{1+2 n}+ \notag \\ &&
2 \sqrt{\pi } \alpha ^2 \xi  \Big(\frac{P_{\text{rc}}}
{\alpha }\Big){}^{3/2} \Big(1+2 n \alpha _1\Big) \theta ^{2+2 n}+2 n \sqrt{\pi
} \alpha ^3 \xi  \Big(\frac{P_{\text{rc}}}{\alpha }
\Big){}^{3/2} \theta ^{3+2 n}-\notag \\ &&
\alpha  P_{\text{rc}} \theta ^{2+n} \frac{d v}
{d \xi}+n P_{\text{rc}} \alpha _1 (-1+v) \theta ^n
\frac{d \theta}{d \xi}-\notag \\ &&
 P_{\text{rc}} \theta ^{1+n} \Big(\alpha _1 \frac{d v}{d \xi}-(1+n)
 \alpha  (-1+v)\frac{d \theta}{d \xi}\Big)\Big)+\notag \\ &&
\sqrt{\pi } E^2 \Big(128 \pi ^{3/2} P_{\text{rc}}^2 \theta ^{1+n}
\Big(\alpha _1+\alpha  \theta \Big)+\alpha ^2 \xi  \sqrt{\frac{P_{\text{rc}}}{\alpha
}} \theta  \Big(1+32 \pi  \Delta -v-\frac{d v}{d \xi}\Big)- \notag \\ &&
2 \sqrt{\pi } \alpha  P_{\text{rc}} \Big(2 (-1+v) \theta +\alpha  \xi
 \Big(-n \xi +4 \sqrt{\pi } \sqrt{\frac{P_{\text{rc}}}{\alpha
}}\Big) \theta ^{2+n}+ \notag \\ &&
16 n \sqrt{\pi } \xi  \sqrt{\frac{P_{\text{rc}}}{\alpha }}
 \alpha _1 \theta ^n \frac{d \theta}{d \xi}-\notag \\ &&
\xi  \theta ^{1+n} (\xi +\Big(n \xi -4 \sqrt{\pi }
\sqrt{\frac{P_{\text{rc}}}{\alpha }}\Big) \alpha _1-16 (1+n) \sqrt{\pi
} \alpha  \sqrt{\frac{P_{\text{rc}}}{\alpha }}
\frac{d \theta}{d \xi}\Bigg)\Bigg)\Bigg)\Bigg)d\xi.
\end{eqnarray}
}
Similarly for case \textbf{2}, Whittaker formula yields
\begin{eqnarray}\label{39}\notag
&&m_L =\int_0^{1}-\frac{\pi\sqrt{(-\frac{2M}{R}+\frac{Q^2}{R^2})
\frac{P_{\text{rc}}}{\alpha }}}{\alpha  \Big(1+\frac{4 \pi ^{3/2}
\xi  E^2}{\sqrt{\frac{P_{\text{rc}}}{\alpha }}}-v\Big)}\notag \\ &&
\Big(-\frac{16 \pi ^{5/2} \alpha  \xi  E^3}{\sqrt{\frac{P_{\text{rc}}}
{\alpha }}}+\frac{\pi ^2 \alpha ^2 \xi  E^4 \Big(3 \xi +32
\sqrt{\pi } \sqrt{\frac{P_{\text{rc}}}{\alpha }}\Big)}{P_{\text{rc}}}
+4 \pi  \alpha  E (-1+v)+ \notag \\ &&
1\Big/\Big(\sqrt{\frac{P_{\text{rc}}}{\alpha }}\Big) \sqrt{\pi }
E^2 \Big(\alpha  \Big(\xi +32 \pi  \Delta  \xi +4 \sqrt{\pi }
\sqrt{\frac{P_{\text{rc}}}{\alpha }}\Big)-\alpha  \Big(\xi +4 \sqrt{\pi }
\sqrt{\frac{P_{\text{rc}}}{\alpha }}\Big) v+ \notag \\ &&
2 \sqrt{\pi } \Big(P_{\text{rc}} \Big(-4 \sqrt{\pi } \xi +64 \pi
 \sqrt{\frac{P_{\text{rc}}}{\alpha }}\Big) \alpha _1+
 \alpha  \xi ^2 \sqrt{\frac{P_{\text{rc}}}{\alpha
}} \Big(1+n \alpha _1\Big)\Big) \theta ^n+\notag \\ &&
 2 \sqrt{\pi } \alpha  \Big(-4 \sqrt{\pi } \xi  P_{\text{rc}}+n
 \alpha  \xi ^2 \sqrt{\frac{P_{\text{rc}}}{\alpha }}+64 \pi
  \alpha  \Big(\frac{P_{\text{rc}}}{\alpha
}\Big){}^{3/2}\Big) \theta ^{1+n}\Big)+\notag \\ &&
 8 \alpha  \Big(\Delta -\Delta  v+2 \sqrt{\pi } \xi
 \Big(\frac{P_{\text{rc}}}{\alpha }\Big){}^{3/2} \theta ^{2 n} \Big(\alpha
_1+\alpha  \theta \Big) \Big(1+n \alpha _1+n \alpha  \theta \Big)\Big)\Big)d\xi.
\end{eqnarray}
\begin{figure}
\centering
\includegraphics[width=80mm]{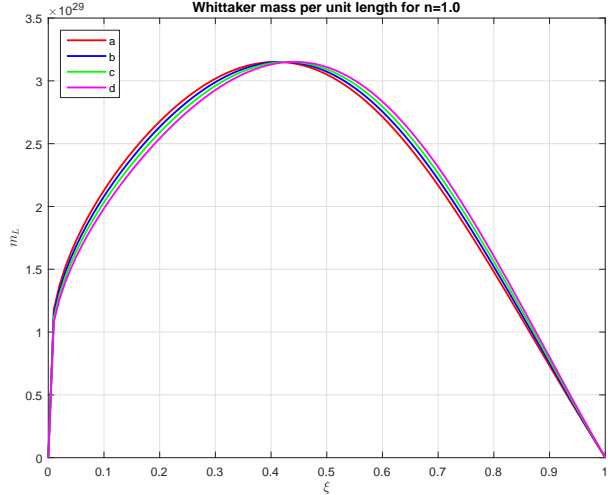}
\caption{Case $1$: $m_L$ as a function of $\xi$ for $n=1$,
curve $a$: $\alpha=8\times 10^{-11}$,~~Q=0.2 $M_\odot$,
curve $b$: $\alpha=10^{-10}$,~~Q=0.4 $M_\odot$,
curve $c$: $\alpha=2 \times 10^{-10}$,~~Q=0.6 $M_\odot$,
curve $d$: $\alpha=4 \times 10^{-10}$,~~Q=0.64 $M_\odot$.}.
\end{figure}
In order to test the physical viability of polytropes, we formulate the energy condition and
found that these conditions are satisfied only for case \textbf{1} and fail in
case \textbf{2} for charged anisotropic
polytropes with GPEoS. We also plot Whittaker formula $(m_L)$ given in Eq. (\ref{38}) with
respect to dimensionless radius $``\xi''$ in radial direction for
a cylinder of unit radius. The bounded behavior of $(m_L)$ in radial direction for various values
of charge shows that our model is stable.
So, only case \textbf{1} for polytropes are physically viable.

\section{Conclusion and Discussion}

In this article, we have formulated the general framework to study
charged relativistic polytropes with GPEoS in cylindrical regime.
The GPEoS $P_r=\alpha_1 \rho +K \rho^{1+\frac{1}{n}}$, is the combination of
linear EoS with polytropic EoS and used in cosmology
for the description of eras of universe.
For the discussion of relativistic polytropes, we have developed
the generalized framework to get the expressions of modified LEE
whose solutions are called polytropes.
The polytropes mainly depends on the density function and the
polytropic index decides the order of solution. These polytropes are very useful
in the description of various astrophysical aspects of CO due
to its simplicity. However, this simplicity is obtained at the cost
of empirical power law relation between density and pressure, which
should hold throughout in CO. The LEE have been obtained with
anisotropic factor $``\Delta''$ in the expressions
(see appendix Eqs. $(\ref{501})$ and $(\ref{502})$) depending upon three variables.
In order to reduce one variable ($``\Delta''$), we have used
conformally flat condition and calculated the value of anisotropic
factor in the form of other two variables (see appendix Eqs. $(\ref{503})$ and $(\ref{504})$).
One can easily obtain modified LE in two variables by substituting these
values of anisotropy factor.

The stability analysis is very important in the development of
mathematical models to check the physical viability.
The energy conditions are very help full in this regard as they can grasp
maximum information without considering EoS involved in the
development of model. The energy conditions have been obtained for
both cases of GPEoS in the presence of electromagnetic field.
We have used modified form of Whittaker formula \cite{35} for the stability
analysis of charged anisotropic relativistic polytropes.
We have calculated $m_L$ which is the measure  of
``active gravitational mass per unit length" in z-direction,
for both case of GPEoS for
different values of parameters involve in model (see Figure \textbf{1})
for cylinder of unit radius. We have taken these values from
our previous work of polytropes \cite{31}.
We have plotted $m_L$ against dimension less radii
$\xi$ and found that its graph remain bounded in radial direction
even for the very high values of charge
$Q=0.2$ $M_\odot$, $0.4$ $M_\odot$, $0.6$ $M_\odot$ and $0.64$ $M_\odot$.
Also it is clear that
more mass is concentrated near the center of cylinder and its
magnitude decreases as we move away from center in radial direction.
The highest magnitude is observed in the middle of cylindrical symmetry
in radial direction. As the gravitational mass cannot be negative
so Figure \textbf{1} shows the stable positive and bounded behavior of
$m_L$, which show that our model is stable.
The energy conditions are valid only for case \textbf{1} and fail to
hold for case \textbf{2}. Hence model developed in case
\textbf{2} is not physically viable due to negation of energy conditions.

\section*{Appendix}

LEE for case \textbf{1}
{
\allowdisplaybreaks
\begin{eqnarray}\label{501}\notag
&&\frac{1}{4} \Big(\frac{\alpha  \Big(\pi  E^2+2 \alpha
 P_{\text{rc}} \theta ^n \Big(\alpha _1+\alpha  \theta \Big)\Big)}{P_{\text{rc}}
\Big((1+n) \alpha  \theta +n \alpha _1 \theta ^{-n}\Big)}-\notag
\\&&
\frac{4 \Big(1+4 \pi ^{3/2} \xi  E^2 \sqrt{\frac{\alpha }
{P_{\text{rc}}}}-v\Big) \theta ^{-n} \Big(8 \pi  E^2+\Delta
+\pi  \xi  \frac{d E}{d \xi} \Big)}{\alpha  \xi ^2 P_{\text{rc}}
 \Big(1+(1+n) \Big(\alpha _1+\alpha  \theta \Big)\Big){}^2}+\notag \\ &&
\Big(2 \theta ^{-n} \Big(8 \pi  E^2+\Delta +\pi
\xi  \frac{d E}{d \xi}\Big) \notag \\ &&
\Big(\frac{8 \pi ^{3/2} \alpha  E^2}{\sqrt{\frac{\alpha }
{P_{\text{rc}}}}}-2 \xi  \Big(P_{\text{rc}} \theta ^n \Big(1+n \alpha _1+n
\alpha  \theta \Big)+\alpha  \xi  \frac{d E}{d \xi}\Big)+\notag \\ &&
\alpha  \xi  E \Big(-4+\pi +\frac{16 \pi ^{3/2} \frac{d E}{d \xi}}
{\sqrt{\frac{\alpha }{P_{\text{rc}}}}}\Big)\Big)\Big)/\Big(\alpha
 \xi  P_{\text{rc}}^2 \Big(1+(1+n) \alpha _1+(1+n) \alpha  \theta
 \Big){}^2\Big)-\notag \\ &&
\frac{\alpha  \xi  \theta ^{-1+n} \Big(-n^2 \alpha _1+(1+n) \alpha
 \theta ^{1+n}\Big) \Big(\pi  E^2+2 \alpha  P_{\text{rc}}
\theta ^n \Big(\alpha _1+\alpha  \theta \Big)\Big) \frac{d \theta}
{d \xi}}{P_{\text{rc}} \Big(n \alpha _1+(1+n) \alpha  \theta
^{1+n}\Big){}^2}-\notag \\ &&
\frac{8 (1+n) \Big(1+4 \pi ^{3/2} \xi  E^2 \sqrt{\frac{\alpha }
{P_{\text{rc}}}}-v\Big) \theta ^{-n} \Big(8 \pi  E^2+\Delta
+\pi  \xi  \frac{d E}{d \xi}\Big) \frac{d \theta}{d \xi}}{\xi
P_{\text{rc}} \Big(1+(1+n) \Big(\alpha _1+\alpha  \theta \Big)\Big){}^3}-\notag \\ &&
\frac{4 n \Big(1+4 \pi ^{3/2} \xi  E^2 \sqrt{\frac{\alpha }
{P_{\text{rc}}}}-v\Big) \theta ^{-1-n} \Big(8 \pi  E^2+\Delta
+\pi  \xi  \frac{d E}{d \xi}\Big)\frac{d \theta}{d \xi}}{\alpha
\xi  P_{\text{rc}} \Big(1+(1+n) \Big(\alpha _1+\alpha  \theta
\Big)\Big){}^2}+\notag \\ &&
\Big(2 \Big(\frac{8 \pi ^{3/2} \alpha  E^2}{\sqrt{\frac{\alpha }
{P_{\text{rc}}}}}-2 \xi  \Big(P_{\text{rc}} \theta ^n \Big(1+n \alpha
_1+n \alpha  \theta \Big)+\alpha  \xi  \frac{d E}{d \xi}\Big)+\notag \\ &&
  \alpha  \xi  E \Big(-4+\pi +\frac{16 \pi ^{3/2} \frac{d E}{d \xi}}
  {\sqrt{\frac{\alpha }{P_{\text{rc}}}}}\Big)\Big) \frac{d \theta}{d \xi}
\Big)/\Big(P_{\text{rc}}
\Big(1+(1+n) \alpha _1+(1+n) \alpha  \theta \Big)\Big)-\notag \\ &&
\frac{4 (1+n) \alpha  \Big(1+4 \pi ^{3/2} \xi  E^2 \sqrt{\frac{\alpha }
{P_{\text{rc}}}}-v\Big) (\frac{d \theta}{d \xi})^2}{\Big(1+(1+n) \Big(\alpha
_1+\alpha  \theta \Big)\Big){}^2}+\notag \\ &&
\frac{2 \alpha  \xi  \theta ^{-1+n} \Big(\pi  E \theta  \frac{d E}
{d \xi}+\alpha  P_{\text{rc}} \theta ^n \Big(n \alpha _1+(1+n)
\alpha  \theta \Big) \frac{d \theta}{d \xi}\Big)}{P_{\text{rc}}
\Big(n \alpha _1+(1+n) \alpha  \theta ^{1+n}\Big)}+\notag \\ &&
\frac{4 \Big(1+4 \pi ^{3/2} \xi  E^2 \sqrt{\frac{\alpha }{P_{\text{rc}}}}
-v\Big) \theta ^{-n} \Big((\pi +16 \pi  E) \frac{d E}{d \xi}
+\frac{d \Delta}{d \xi}+\pi  \xi  \frac{d^2 E}{d \xi^2} \Big)}{\alpha  \xi
 P_{\text{rc}} \Big(1+(1+n) \alpha _1+(1+n) \alpha  \theta \Big){}^2}+\notag \\ &&
 \frac{4 \Big(1+4 \pi ^{3/2} \xi  E^2 \sqrt{\frac{\alpha }{P_{\text{rc}}}}
 -v\Big) \frac{d^2 \theta}{d \xi^2}}{1+(1+n) \Big(\alpha _1+\alpha
 \theta \Big)}\Big)=0.
\end{eqnarray}
}
LEE for case \textbf{2}
{
\allowdisplaybreaks
\begin{eqnarray}\label{502}\notag
&&\frac{1}{4} \Big(\frac{\alpha  \Big(\pi  E^2+2 \alpha  P_{\text{rc}}
 \theta ^n \Big(\alpha _1+\alpha  \theta
 \Big)\Big)}{P_{\text{rc}}
\Big((1+n) \alpha  \theta +n \alpha _1 \theta ^{-n}\Big)}-\notag \\ &&
\frac{4 \Big(1+4 \pi ^{3/2} \xi  E^2 \sqrt{\frac{\alpha }{P_{\text{rc}}}}
-v\Big) \theta ^{-n} \Big(8 \pi  E^2+\Delta
+\pi  \xi  \frac{d E}{d \xi}\Big)}{\alpha  \xi ^2 P_{\text{rc}} \Big(1
+\alpha _1+\alpha  \theta \Big) \Big(1+(1+n) \Big(\alpha _1+\alpha
 \theta \Big)\Big)}+\notag \\ &&
\Big(2 \theta ^{-n} \Big(8 \pi  E^2+\Delta +\pi  \xi  \frac{d E}{d \xi}
\Big) .\notag \\ &&
\Big(\frac{8 \pi ^{3/2} \alpha  E^2}{\sqrt{\frac{\alpha }{P_{\text{rc}}}}}
-2 \xi  \Big(P_{\text{rc}} \theta ^n+\alpha
\xi  \frac{d E}{d \xi}\Big)+\alpha  \xi  E \Big(-4+\pi +\frac{16 \pi ^{3/2}
 \frac{d E}{d \xi}}{\sqrt{\frac{\alpha }{P_{\text{rc}}}}}\Big)\Big)\Big)
 \Big/\notag \\ &&
\Big(\alpha  \xi  P_{\text{rc}}^2 \Big(1+\alpha _1+\alpha  \theta \Big)
\Big(1+(1+n) \alpha _1+(1+n) \alpha  \theta \Big)\Big)-\notag \\ &&
\frac{\alpha  \xi  \theta ^{-1+n} \Big(-n^2 \alpha _1+(1+n) \alpha
 \theta ^{1+n}\Big) \Big(\pi  E^2+2 \alpha  P_{\text{rc}}
\theta ^n \Big(\alpha _1+\alpha  \theta \Big)\Big) \frac{d \theta}
{d \xi}}{P_{\text{rc}} \Big(n \alpha _1+(1+n) \alpha  \theta
^{1+n}\Big){}^2}-\notag \\ &&
\frac{4 (1+n) \Big(1+4 \pi ^{3/2} \xi  E^2 \sqrt{\frac{\alpha }
{P_{\text{rc}}}}-v\Big) \theta ^{-n} \Big(8 \pi  E^2+\Delta
+\pi  \xi  \frac{d E}{d \xi}\Big) \frac{d \theta}{d \xi}}{\xi
 P_{\text{rc}} \Big(1+\alpha _1+\alpha  \theta \Big) \Big(1+(1+n) \Big(\alpha _1+\alpha
 \theta \Big)\Big){}^2}-\notag \\ &&
\frac{4 \Big(1+4 \pi ^{3/2} \xi  E^2 \sqrt{\frac{\alpha }
{P_{\text{rc}}}}-v\Big) \theta ^{-n} \Big(8 \pi  E^2+\Delta
+\pi  \xi  \frac{d E}{d \xi}\Big) \frac{d \theta}{d \xi}}
{\xi  P_{\text{rc}} \Big(1+\alpha _1+\alpha  \theta \Big){}^2 \Big(1+(1+n) \Big(\alpha
_1+\alpha  \theta \Big)\Big)}-\notag \\ &&
\frac{4 n \Big(1+4 \pi ^{3/2} \xi  E^2 \sqrt{\frac{\alpha }
{P_{\text{rc}}}}-v\Big) \theta ^{-1-n} \Big(8 \pi  E^2+\Delta
+\pi  \xi  \frac{d E}{d \xi}\Big)\frac{d \theta}{d \xi}}
{\alpha  \xi  P_{\text{rc}} \Big(1+\alpha _1+\alpha
 \theta \Big) \Big(1+(1+n) \Big(\alpha
_1+\alpha  \theta \Big)\Big)}+\notag \\ &&
\frac{2 \Big(\frac{8 \pi ^{3/2} \alpha  E^2}{\sqrt{
\frac{\alpha }{P_{\text{rc}}}}}-2 \xi  \Big(P_{\text{rc}}
\theta ^n+\alpha  \xi
\frac{d E}{d \xi}\Big)+\alpha  \xi  E \Big(-4+\pi +
\frac{16 \pi ^{3/2} \frac{d E}{d \xi}}{\sqrt{\frac{\alpha }{P_{\text{rc}}}}}\Big)\Big)
\frac{d \theta}{d \xi}}{P_{\text{rc}} \Big(1+\alpha _1+\alpha  \theta \Big)}-\notag \\ &&
\frac{4 \alpha  \Big(1+4 \pi ^{3/2} \xi  E^2 \sqrt{\frac{\alpha }{P_{\text{rc}}}}-v\Big)
 (\frac{d \theta}{d \xi})^2}{\Big(1+\alpha _1+\alpha
 \theta \Big){}^2}+\notag \\ &&
\frac{2 \alpha  \xi  \theta ^{-1+n} \Big(\pi  E \theta  \frac{d E}
{d \xi}+\alpha  P_{\text{rc}} \theta ^n \Big(n \alpha _1+(1+n)
\alpha  \theta \Big) \frac{d \theta}{d \xi}\Big)}{P_{\text{rc}}
 \Big(n \alpha _1+(1+n) \alpha  \theta ^{1+n}\Big)}+\notag \\ &&
\frac{4 \Big(1+4 \pi ^{3/2} \xi  E^2 \sqrt{\frac{\alpha }{P_{\text{rc}}}}
-v\Big) \theta ^{-n} \Big((\pi +16 \pi  E)
\frac{d E}{d \xi}+\frac{d \Delta}{d \xi}+\pi  \xi \frac{d^2 E}{d \xi}\Big)}
{\alpha  \xi  P_{\text{rc}} \Big(1+\alpha _1+\alpha  \theta \Big) \Big(1+(1+n) \alpha _1+(1+n)
\alpha  \theta \Big)}+\notag \\ &&
 \frac{4 \Big(1+4 \pi ^{3/2} \xi  E^2 \sqrt{\frac{\alpha }{P_{\text{rc}}}}
 -v\Big) \frac{d^2 \theta}{d \xi^2}}{1+\alpha _1+\alpha  \theta
}\Big)=0.
\end{eqnarray}
}
$\Delta$ For Case \textbf{1}
{
\allowdisplaybreaks
\begin{eqnarray}\label{503}\notag
\Delta &=&-\pi  E^2+\frac{1}{8 \pi }\Big(4 \pi ^2 E^2+\frac{8 \pi
P_{\text{rc}} \theta ^n \Big(\alpha _1+\alpha  \theta
\Big)}{\alpha }+\notag \\ &&
\Big(4 \pi ^2 E^2+\frac{8 \pi  P_{\text{rc}} \theta ^n \Big(\alpha _1
+\alpha  \theta \Big)}{\alpha }\Big){}^2+\Big(4 \pi
^2 E^2+\frac{8 \pi  P_{\text{rc}} \theta ^n \Big(\alpha _1+\alpha
\theta \Big)}{\alpha }\Big) \notag \\ &&
\Big(-4 \pi ^2 E^2+\frac{8 \pi  P_{\text{rc}} \theta ^n \Big(1+n
\alpha _1+n \alpha  \theta \Big)}{\alpha }\Big)+\notag \\ &&
\Big(-4 \pi ^2 E^2+\frac{8 \pi  P_{\text{rc}} \theta ^n \Big(1+n
\alpha _1+n \alpha  \theta \Big)}{\alpha }\Big){}^2+\notag \\ &&
\Big(-1-8 \pi ^2 E^2-\frac{8 \pi  P_{\text{rc}} \theta ^n \Big
(\alpha _1+\alpha  \theta \Big)}{\alpha }+\frac{8 \pi  P_{\text{rc}}
\theta ^n \Big(1+n \alpha _1+n \alpha  \theta \Big)}{\alpha }\Big) \notag \\ &&
\Big(\frac{\Big(\frac{\pi ^{3/2} \xi  E^2}{\sqrt{\frac{P_{\text{rc}}}
{\alpha }}}+\frac{8 \pi  P_{\text{rc}} \theta ^n \Big(\alpha
_1+\alpha  \theta \Big)}{\alpha }\Big){}^2}{\Big(1+\frac{4 \pi ^{3/2}
 \xi  E^2}{\sqrt{\frac{P_{\text{rc}}}{\alpha }}}-
v\Big){}^2}-\notag \\ &&
\frac{\xi  \Big(\frac{\pi ^{3/2} \xi  E^2}{\sqrt{\frac{P_{\text{rc}}}
{\alpha }}}+\frac{8 \pi  P_{\text{rc}} \theta ^n \Big(\alpha _1+\alpha
 \theta \Big)}{\alpha }\Big) \Big(-4 \pi ^2 E^2+\frac{8 \pi
 P_{\text{rc}} \theta ^n \Big(1+n \alpha _1+n \alpha  \theta
\Big)}{\alpha }\Big)}{4 \sqrt{\pi } \sqrt{\frac{P_{\text{rc}}}{\alpha }}
 \Big(1+\frac{4 \pi ^{3/2} \xi  E^2}{\sqrt{\frac{P_{\text{rc}}}{\alpha
}}}-v\Big)}-\notag \\ &&
\frac{\Big(\frac{\pi ^{3/2} \xi  E^2}{\sqrt{\frac{P_{\text{rc}}}{\alpha }}}
+\frac{8 \pi  P_{\text{rc}} \theta ^n \Big(\alpha _1+\alpha
 \theta \Big)}{\alpha }\Big) \Big(16 \pi ^2 E^2+32 \pi ^2 \xi  E
 \frac{d E}{d \xi}-\frac{d v}{d \xi}\Big)}{\Big(1+\frac{4 \pi ^{3/2} \xi
 E^2}{\sqrt{\frac{P_{\text{rc}}}{\alpha }}}-v\Big){}^2}+\notag \\ &&
 \frac{4 \pi ^2 E^2+8 \pi ^2 \xi  E \frac{d E}{d \xi}+\frac{8 \pi
 P_{\text{rc}} \theta ^{-1+n} \Big(n \alpha _1+(1+n) \alpha  \theta
\Big) \frac{d \theta}{d \xi}}{\alpha }}{1+\frac{4 \pi ^{3/2} \xi
E^2}{\sqrt{\frac{P_{\text{rc}}}{\alpha }}}-v}\Big)+\notag \\ &&
\Big(\frac{\Big(\frac{\pi ^{3/2} \xi  E^2}{\sqrt{\frac{P_{\text{rc}}}
{\alpha }}}+\frac{8 \pi  P_{\text{rc}} \theta ^n \Big(\alpha
_1+\alpha  \theta \Big)}{\alpha }\Big){}^2}{\Big(1+\frac{4 \pi ^{3/2}
 \xi  E^2}{\sqrt{\frac{P_{\text{rc}}}{\alpha }}}-
v\Big){}^2}-\notag \\ &&
\frac{\xi  \Big(\frac{\pi ^{3/2} \xi  E^2}{\sqrt{\frac{P_{\text{rc}}}
{\alpha }}}+\frac{8 \pi  P_{\text{rc}} \theta ^n \Big(\alpha _1+\alpha
 \theta \Big)}{\alpha }\Big) \Big(-4 \pi ^2 E^2+\frac{8 \pi
 P_{\text{rc}} \theta ^n \Big(1+n \alpha _1+n \alpha  \theta
\Big)}{\alpha }\Big)}{4 \sqrt{\pi } \sqrt{\frac{P_{\text{rc}}}{\alpha }}
 \Big(1+\frac{4 \pi ^{3/2} \xi  E^2}{\sqrt{\frac{P_{\text{rc}}}{\alpha
}}}-v\Big)}-\notag \\ &&
\frac{\Big(\frac{\pi ^{3/2} \xi  E^2}{\sqrt{\frac{P_{\text{rc}}}{\alpha }}}
+\frac{8 \pi  P_{\text{rc}} \theta ^n \Big(\alpha _1+\alpha
 \theta \Big)}{\alpha }\Big) \Big(16 \pi ^2 E^2+32 \pi ^2 \xi  E
 \frac{d E}{d \xi}-\frac{d v}{d \xi}\Big)}{\Big(1+\frac{4 \pi ^{3/2} \xi
 E^2}{\sqrt{\frac{P_{\text{rc}}}{\alpha }}}-v\Big){}^2}+\notag \\ &&
  \frac{4 \pi ^2 E^2+8 \pi ^2 \xi  E \frac{d E}{d \xi}+\frac{8 \pi
   P_{\text{rc}} \theta ^{-1+n} \Big(n \alpha _1+(1+n) \alpha
 \theta \Big) \frac{d \theta}{d \xi}}{\alpha }}{1+\frac{4 \pi ^{3/2}
  \xi  E^2}{\sqrt{\frac{P_{\text{rc}}}{\alpha }}}-v}\Big){}^2\Big).
\end{eqnarray}
}
$\Delta$ For Case \textbf{2}
{
\allowdisplaybreaks
\begin{eqnarray}\label{504}\notag
\Delta &=&-\pi  E^2+\frac{1}{8 \pi }\Big(4 \pi ^2 E^2+\frac{8 \pi
P_{\text{rc}} \theta ^n \Big(\alpha _1+\alpha
\theta \Big)}{\alpha }+\Big(-4 \pi ^2 E^2+\frac{8 \pi  P_{\text{rc}}
 \theta ^n}{\alpha }\Big){}^2+\notag \\ &&
\Big(-4 \pi ^2 E^2+\frac{8 \pi  P_{\text{rc}} \theta ^n}{\alpha }\Big)
 \Big(4 \pi ^2 E^2+\frac{8 \pi  P_{\text{rc}} \theta
^n \Big(\alpha _1+\alpha  \theta \Big)}{\alpha }\Big)+\notag \\ &&
\Big(4 \pi ^2 E^2+\frac{8 \pi  P_{\text{rc}} \theta ^n \Big(\alpha _1
+\alpha  \theta \Big)}{\alpha }\Big){}^2+\notag \\ &&
\Big(-1-8 \pi ^2 E^2+\frac{8 \pi  P_{\text{rc}} \theta ^n}{\alpha }
-\frac{8 \pi  P_{\text{rc}} \theta ^n \Big(\alpha _1+\alpha
 \theta \Big)}{\alpha }\Big) \notag \\ &&
\Big(-\frac{\xi  \Big(-4 \pi ^2 E^2+\frac{8 \pi  P_{\text{rc}}
\theta ^n}{\alpha }\Big) \Big(\frac{\pi ^{3/2} \xi
  E^2}{\sqrt{\frac{P_{\text{rc}}}{\alpha
}}}+\frac{8 \pi  P_{\text{rc}} \theta ^n \Big(\alpha _1+\alpha
\theta \Big)}{\alpha }\Big)}{4 \sqrt{\pi } \sqrt{\frac{P_{\text{rc}}}{\alpha
}} \Big(1+\frac{4 \pi ^{3/2} \xi  E^2}{\sqrt{\frac{P_{\text{rc}}}
{\alpha }}}-v\Big)}+\notag \\ &&
\frac{\Big(\frac{\pi ^{3/2} \xi  E^2}{\sqrt{\frac{P_{\text{rc}}}
{\alpha }}}+\frac{8 \pi  P_{\text{rc}} \theta ^n \Big(\alpha _1+\alpha
 \theta \Big)}{\alpha }\Big){}^2}{\Big(1+\frac{4 \pi ^{3/2} \xi
 E^2}{\sqrt{\frac{P_{\text{rc}}}{\alpha }}}-v\Big){}^2}-\notag \\ &&
\frac{\Big(\frac{\pi ^{3/2} \xi  E^2}{\sqrt{\frac{P_{\text{rc}}}
{\alpha }}}+\frac{8 \pi  P_{\text{rc}} \theta ^n \Big(\alpha _1+\alpha
 \theta \Big)}{\alpha }\Big) \Big(16 \pi ^2 E^2+32 \pi ^2 \xi
 E \frac{d E}{d \xi}-\frac{d v}{d \xi}\Big)}{\Big(1+\frac{4 \pi ^{3/2} \xi
 E^2}{\sqrt{\frac{P_{\text{rc}}}{\alpha }}}-v\Big){}^2}+\notag \\ &&
 \frac{4 \pi ^2 E^2+8 \pi ^2 \xi  E \frac{d E}{d \xi}+\frac{8 \pi
 P_{\text{rc}} \theta ^{-1+n} \Big(n \alpha _1+(1+n) \alpha  \theta
\Big) \frac{d \theta}{d \xi} }{\alpha }}{1+\frac{4 \pi ^{3/2} \xi
E^2}{\sqrt{\frac{P_{\text{rc}}}{\alpha }}}-v}\Big)+\notag \\ &&
\Big(-\frac{\xi  \Big(-4 \pi ^2 E^2+\frac{8 \pi  P_{\text{rc}}
\theta ^n}{\alpha }\Big) \Big(\frac{\pi ^{3/2} \xi
  E^2}{\sqrt{\frac{P_{\text{rc}}}{\alpha
}}}+\frac{8 \pi  P_{\text{rc}} \theta ^n \Big(\alpha _1+\alpha
\theta \Big)}{\alpha }\Big)}{4 \sqrt{\pi } \sqrt{\frac{P_{\text{rc}}}
{\alpha
}} \Big(1+\frac{4 \pi ^{3/2} \xi  E^2}{\sqrt{\frac{P_{\text{rc}}}
{\alpha }}}-v\Big)}+\notag \\ &&
\frac{\Big(\frac{\pi ^{3/2} \xi  E^2}{\sqrt{\frac{P_{\text{rc}}}
{\alpha }}}+\frac{8 \pi  P_{\text{rc}} \theta ^n \Big(\alpha _1+\alpha
 \theta \Big)}{\alpha }\Big){}^2}{\Big(1+\frac{4 \pi ^{3/2} \xi  E^2}
 {\sqrt{\frac{P_{\text{rc}}}{\alpha }}}-v\Big){}^2}-\notag \\ &&
\frac{\Big(\frac{\pi ^{3/2} \xi  E^2}{\sqrt{\frac{P_{\text{rc}}}
{\alpha }}}+\frac{8 \pi  P_{\text{rc}} \theta ^n \Big(\alpha _1+\alpha
 \theta \Big)}{\alpha }\Big) \Big(16 \pi ^2 E^2+32 \pi ^2 \xi
 E \frac{d E}{d \xi}-\frac{d v}{d \xi}\Big)}{\Big(1+\frac{4 \pi ^{3/2} \xi
 E^2}{\sqrt{\frac{P_{\text{rc}}}{\alpha }}}-v\Big){}^2}+\notag \\ &&
  \frac{4 \pi ^2 E^2+8 \pi ^2 \xi  E \frac{d E}{d \xi}+\frac{8 \pi
  P_{\text{rc}} \theta ^{-1+n} \Big(n \alpha _1+(1+n) \alpha
 \theta \Big) \frac{d \theta}{d \xi}}{\alpha }}{1+\frac{4 \pi ^{3/2}
  \xi  E^2}{\sqrt{\frac{P_{\text{rc}}}{\alpha }}}-v}\Big){}^2\Big).
\end{eqnarray}
}

\end{document}